\documentclass[twocolumn,showpacs,showkeys,amsmath,amssymb,aps,pre,superscriptaddress]{revtex4}
\usepackage{graphicx}
\usepackage{dcolumn}
\usepackage{bm}

\begin{document}
\title{Spin-1/2 Ising-Heisenberg model with the pair XYZ Heisenberg interaction 
and quartic Ising interactions as the exactly soluble zero-field eight-vertex model}
\author{Jozef Stre\v{c}ka}
\email{jozef.strecka@upjs.sk, jozkos@pobox.sk}
\affiliation{Department of Theoretical Physics and Astrophysics, 
Faculty of Science, \\ P. J. \v{S}af\'{a}rik University, Park Angelinum 9,
040 01 Ko\v{s}ice, Slovak Republic}
\author{Lucia \v{C}anov\'a}
\affiliation{Department of Applied Mathematics, Faculty of Mechanical Engineering, \\ 
Technical University, Letn\'a 9, 042 00 Ko\v{s}ice, Slovak Republic}
\author{Kazuhiko Minami}
\affiliation{Graduate School of Mathematics, Nagoya University,                  
             Nagoya 464-8602, Japan}      
\date{\today}
\begin{abstract}
The spin-1/2 Ising-Heisenberg model with the pair XYZ Heisenberg interaction and 
quartic Ising interactions is exactly solved by establishing a precise mapping 
relationship with the corresponding zero-field (symmetric) eight-vertex model. It is shown 
that the Ising-Heisenberg model with the ferromagnetic Heisenberg interaction exhibits 
a striking critical behavior, which manifests itself through re-entrant phase transitions 
as well as continuously varying critical exponents. The changes of critical exponents 
are in accordance with the weak universality hypothesis in spite of a peculiar singular 
behavior to emerge at a quantum critical point of the infinite order, which occurs at 
the isotropic limit of the Heisenberg interaction. 
On the other hand, the Ising-Heisenberg model with the antiferromagnetic Heisenberg 
interaction surprisingly exhibits less significant changes of both critical temperatures 
as well as critical exponents upon varying a strength of the exchange anisotropy in 
the Heisenberg interaction.
\end{abstract}


\maketitle

\section{Introduction}
\label{sec1}

Over the last six decades, exactly solvable models of interacting many-body systems have 
attracted considerable attention due to their uncoverable role in a fundamental understanding 
of order-disorder phenomena \cite{baxt82,diep04,lavi99}. In particular, exactly solvable quantum 
spin models are currently at a forefront of theoretical research interest as they often exhibit 
a peculiar interplay between quantum fluctuations and cooperativity \cite{matt93,lieb04,suth04}. 
Despite extensive studies, however, phase transitions and critical phenomena of rigorously solved 
quantum spin models still belong to the most challenging unresolved issues to deal with, 
since quantum fluctuations usually prefer a lack of spontaneous order. Indeed, there are 
only few exactly solved quantum spin models that simultaneously exhibit both spontaneous 
long-range order, as well as, obvious quantum manifestations. 

It has been recently demonstrated that the spontaneous order, which is accompanied with 
obvious quantum manifestations, might be an inherent feature of the hybrid Ising-Heisenberg 
planar models \footnote{The term Ising-Heisenberg model might be a little bit confusing 
as it closely resembles the term Heisenberg-Ising model, which is often used to refer 
to the XXZ Heisenberg models with the Ising-like anisotropy. Unlike this case, the term Ising-Heisenberg model will be strictly used in connection with hybrid models that 
consist of both Ising as well as Heisenberg spins.} whose lattice sites are in part 
occupied by the Ising spins and partly by the Heisenberg spins \cite{stre02,jasc02,stre04,jasc04,stre06,cano07,stre08,yao08,cano08,jasc08}. 
Note that all aforementioned Ising-Heisenberg planar models
have exactly been solved with the help of suitable algebraic transformations such as 
the decoration-iteration \cite{syoz51} or the star-triangle \cite{onsa44} transformation, 
which establish a precise mapping equivalence with the corresponding spin-1/2 Ising model. 
It is worthwhile to recall, moreover, that the decoration-iteration and star-triangle mapping transformations were substantially generalized by Fisher \cite{fish59} (see for details also Ref.~\cite{syoz72}) who firstly pointed out that in principle arbitrary quantum-mechanical 
system coupled to two or three outer Ising spins can be replaced via appropriate algebraic transformation by the equivalent expression containing pairwise spin-spin interactions 
between the outer Ising spins. In such a way, one effectively establishes a precise
mapping relationship that connects the exact solution of some original model (which might 
describe a rather complex quantum-mechanical system) with the exact solution of the 
corresponding spin-1/2 Ising model, which is generally known for many planar lattices 
of different topologies \cite{baxt82,lavi99,syoz72,utiy51,domb60,lin86}.

On the contrary, there does not exist a general algebraic transformation for any 
quantum-mechanical system coupled to four or more outer Ising spins if one considers 
pairwise interactions between the outer Ising spins only \cite{fish59}. Rojas \textit{et al}. \cite{roja08} has recently found an exact evidence that it is nevertheless possible 
to include multispin interactions between the outer Ising spins into the algebraic 
transformation in order to ensure its general validity. Thus, the algebraic transformation 
with effective pair and quartic interactions is for instance required when searching for 
an exact treatment of the quantum-mechanical system coupled to four outer Ising spins 
in an absence of the external field. The spin-1/2 Ising model with pair and quartic 
interactions is however nothing but the alternative definition of the general eight-vertex model \cite{wu71,kada71}, which becomes exactly tractable by imposing a special additional constraint 
to its vertex energies (Boltzmann's weights) known either as the Baxter's zero-field (symmetric) condition \cite{baxt82,baxt71,baxt72} or the free-fermion condition of Fan and Wu \cite{fan69,fan70}. 
Bearing all this in mind, one could intuitively expect that there might exist a certain 
class exactly solvable quantum-mechanical spin models for which the appropriate algebraic transformation yields a precise mapping correspondence to the eight-vertex model generally 
satisfying the Baxter's zero-field condition \cite{baxt82,baxt71,baxt72}. The main goal 
of the present work is to show that the spin-1/2 Ising-Heisenberg model with the pair 
XYZ Heisenberg interaction and the quartic Ising interactions falls into this class 
of fully exactly soluble models. 

An importance of exact results to be obtained for the Ising-Heisenberg model with the pair 
and quartic interactions should be viewed in more respects. First, it is well known that 
the quartic interaction basically affects magnetic properties of several insulating copper 
compounds such as La$_2$CuO$_4$ \cite{roge89,suga90,hond93,mizu99,cold01,kata02} 
(undoped parent compound for high-$T_c$ cuprates), SrCu$_2$O$_3$ \cite{mizu99}, La$_6$Ca$_8$Cu$_{24}$O$_{41}$ \cite{mats00}, and La$_4$Sr$_{10}$Cu$_{24}$O$_{41}$ \cite{notb07}. 
Even though it would be rather striking coincidence if some real magnetic material 
would obey very specific topological requirements of the model under investigation, 
it is quite reasonable to suspect that our exact results might at least shed light 
on some important aspects of the critical behavior of real magnetic materials. 
Second, our model system represents a rare example of the exactly solved lattice-statistical 
model, which contradicts the standard universality hypothesis in that its critical 
exponents vary continuously with the interaction parameters over the full range 
of possible values of the critical exponents. Third, the exactly soluble vertex models 
have found over the last few decades manifold applications in seemingly diverse research 
areas. Exact solutions of vertex-like models essentially tackles the problem of the 
residual entropy of two-dimensional ice \cite{lieb67a,lieb67b}, 
the Slater model of hydrogen-bonded ferroelectrics \cite{wu67,suth67,lieb67d}, 
the integrable quantum spin models such as the quantum Heisenberg chain \cite{lieb67c,suth67,baxt71b,baxt73a,baxt73b,baxt73c,baxt02}, 
the ice-type solid-on-solid (SOS) model \cite{baxt73b,baxt73c,andr84,bazh07}, 
the problem of counting domino tilings \cite{fan70,kore00,zinn00,cohn00,mina08}, 
the three- and four-coloring problems of the square and hexagonal lattices \cite{baxt82,lieb67b,baxt70a,baxt70b,liebwu}, etc.

The rest of this paper is organized as follows. In the next section, we will provide the detailed description of the hybrid Ising-Heisenberg model and then, basic ideas of the exact mapping 
to the zero-field eight-vertex model will be explained. This is followed by the presentation 
of the most interesting results for the ground-state and finite-temperature phase diagrams, 
which are supplemented by a detailed analysis of a rather strange non-universal behavior 
of critical exponents. Finally, some concluding remarks are given in Section \ref{sec4}.

\section{Ising-Heisenberg model and its equivalence to the zero-field eight-vertex model}
\label{sec2}

\begin{figure}[htb]
\vspace{0cm}
\begin{center}
\includegraphics[width=5cm]{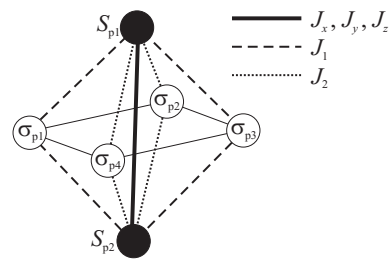}
\end{center}
\vspace{-0.7cm}
\caption{The elementary unit cell of the spin-1/2 Ising-Heisenberg model. Full (empty) circles
denote lattice positions of the Heisenberg (Ising) spins, thick solid line represents the pairwise
XYZ Heisenberg interaction between the apical Heisenberg spins and both types of broken lines connect 
spins involved in the quartic Ising interactions. Thin solid lines connecting four Ising spins 
are guide for eyes only.}
\label{fig1}
\end{figure}

Consider a two-dimensional lattice of edge-sharing octahedrons, each of them having four 
Ising spins $\sigma=1/2$ in a basal plane and two Heisenberg spins $S=1/2$ in apical positions, 
as it is schematically depicted in Fig.~\ref{fig1}. Suppose furthermore 
that each edge of the octahedron, which connects two Ising spins, is a common edge of two 
adjacent octahedrons (thin solid lines in Fig.~\ref{fig1}). The ensemble of all Ising spins 
then forms a square lattice, which has a couple of the Heisenberg spins above and below a center 
of each elementary square face formed by four Ising spins. Let both apical Heisenberg spins 
interact together via the pairwise XYZ Heisenberg interaction,
while they are also assumed to be engaged in two different quartic Ising interactions 
with two Ising spins from opposite corners of a square face (see Fig.~\ref{fig1}). The total Hamiltonian can be for convenience written as a sum over all elementary unit cells (octahedrons) $\hat{{\cal H}} = \sum_{p} \hat{{\cal H}}_p$, 
where each octahedron-cluster Hamiltonian $\hat{{\cal H}}_p$ contains one pair interaction between 
the apical Heisenberg spins and two quartic interactions between the Heisenberg spins 
and their four Ising neighbors 
\begin{eqnarray}
\hat{{\cal H}}_p = 
  \! \! \! &-& \! \! \! \left(J_x \hat{S}_{p1}^x \hat{S}_{p2}^x + J_y \hat{S}_{p1}^y \hat{S}_{p2}^y 
                                  + J_z \hat{S}_{p1}^z \hat{S}_{p2}^z \right) \nonumber \\
  \! \! \! &-& \! \! \! J_1 \hat{S}_{p1}^z \hat{S}_{p2}^z \hat{\sigma}_{p1}^z \hat{\sigma}_{p3}^z 
            - J_2 \hat{S}_{p1}^z \hat{S}_{p2}^z \hat{\sigma}_{p2}^z \hat{\sigma}_{p4}^z. 
\label{ham}	   
\end{eqnarray}
Above, the interaction parameters $J_x, J_y, J_z$ denote spatial components of the anisotropic 
XYZ interaction between the Heisenberg spins, while the interaction parameters $J_1$ and $J_2$ 
label two quartic Ising interactions between both apical Heisenberg spins and two Ising spins 
from opposite corners of a square face along two different diagonal directions 
(see Fig.~\ref{fig1}).

It is of principal importance that the cluster Hamiltonians of two different octahedrons commute 
with each other, i.e. $[\hat{{\cal H}}_i, \hat{{\cal H}}_j] = 0$, which immediately allows a partial factorization of the total partition function into a product of cluster partition functions
\begin{eqnarray}
{\mathcal Z} = \sum_{\{\sigma \}} \prod_{p} \mbox{Tr}_p \exp(- \beta \hat{{\cal H}}_p).    
\label{pf1}
\end{eqnarray}
The summation $\sum_{\{\sigma \}}$ to emerge in Eq.~(\ref{pf1}) is carried out over all possible 
configurations of the Ising spins, the symbol $\mbox{Tr}_p$ denotes a trace over spin degrees
of freedom of the Heisenberg spin pair from the $p$th octahedron, $\beta = 1/(k_{\rm B} T)$, 
$k_{\rm B}$ is Boltzmann's constant and $T$ is the absolute temperature. After performing 
the relevant trace over spin degrees of freedom of the Heisenberg spins, the partition 
function of the Ising-Heisenberg model can be rewritten into the form  
\begin{eqnarray}
{\mathcal Z} = \sum_{\{\sigma \}} \prod_{p} 
\omega_p \left(\sigma^z_{p1}, \sigma^z_{p2}, \sigma^z_{p3}, \sigma^z_{p4} \right).    
\label{pf2}
\end{eqnarray} 
Apparently, the effective Boltzmann's factor $\omega_p$ assigned to the $p$th octahedron  
now explicitly depends just on four Ising spins $\sigma_{p1}$, $\sigma_{p2}$, $\sigma_{p3}$ 
and $\sigma_{p4}$ from its basal plane through the relation
\begin{widetext}
\begin{eqnarray}
\omega_p (a, b, c, d) = 
         2 \exp \left(K_z + K_1 ac + K_2 bd \right)   \cosh \left(K_x - K_y \right)
       + 2 \exp \left(- K_z - K_1 ac - K_2 bd \right) \cosh \left(K_x + K_y \right), 
\label{bf} 
\end{eqnarray}
\end{widetext}
where we have introduced a new unified notation for the coupling constants $K_{\alpha} = \beta J_{\alpha}/4$ ($\alpha=x,y,z,1,2$) in order to write the Boltzmann's factor (\ref{bf}) 
in a more abbreviated and elegant form. 

\begin{figure}[htb]
\begin{center}
\includegraphics[width=8.5cm]{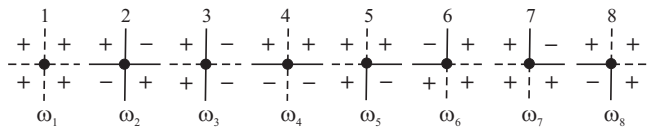}
\end{center}
\vspace{-0.6cm}
\caption{Eight possible Ising spin configurations and their relation to line coverings of 
the corresponding eight-vertex model on a dual square lattice. The sign '$\pm$' denotes 
the spin state $\sigma^z = \pm 1/2$.}
\label{fig2}
\end{figure}

At this stage, the model under investigation can be rather straightforwardly mapped to the eight-vertex model on its dual square lattice. As a matter of fact, the product emerging in Eq.~(\ref{pf2}) can alternatively be performed over all elementary squares of the Ising spins forming a square lattice 
and the Boltzmann's factor (\ref{bf}) is invariant under the reversal of all four Ising spin variables. In this respect, there are at the utmost eight different spin arrangements that have different energies (Boltzmann's weights) and these can readily be related to the Boltzmann's weights 
of the eight-vertex model on a dual square lattice by the following procedure. If, and only if, the Ising spins located at adjacent corners of a square face are aligned opposite to each other, then solid lines are drawn on edges of a dual square lattice, otherwise they are drawn as broken lines. 
This is actually one of many alternative definitions of the eight-vertex model, since an even number 
of solid (broken) lines is always incident to each vertex of a dual square lattice. The diagrammatic representation of eight possible spin arrangements and their corresponding line coverings is shown in Fig.~\ref{fig2}.  
It can be easily understood that each of eight possible line arrangements around a vertex of 
the dual lattice corresponds to two spin configurations, one is being obtained from the other 
by reversing all four Ising spins located at corners of an elementary square face. The hybrid Ising-Heisenberg model thus turns out to be equivalent with the eight-vertex model. With regard 
to this equivalence, the partition function of the Ising-Heisenberg model can be expressed 
in terms of the partition function of the eight-vertex model on a square lattice
\begin{eqnarray}
{\mathcal Z} (T, J_x, J_y, J_z, J_1, J_2) 
= 2 {\mathcal Z}_{8-{\rm vertex}} (\omega_1, \omega_2, ..., \omega_8).
\label{PF}
\end{eqnarray}   
The factor 2 in the above equation comes from the two-to-one mapping between the spin and vertex 
configurations (two spin configurations always correspond to one vertex configuration). 

The Boltzmann's weights, which correspond to eight possible line coverings of the eight-vertex model 
shown in Fig. \ref{fig2}, can directly be obtained from Eq.~(\ref{bf}) by substituting 
a respective spin configuration of the Ising spins 
\begin{eqnarray}
\omega_1 \! \! &=& \! \! \omega_2 =
2 \exp \left(K_z + \frac{K_1 + K_2}{4} \right) \cosh \left(K_x - K_y \right) \nonumber \\
       \! \! \!&+& \! \! \! 2 \exp \left(- K_z - \frac{K_1 + K_2}{4} \right)  
       \cosh \left(K_x + K_y \right)\!, \label{bw1} \\
\omega_3 \! \! &=& \! \! \omega_4 =
2 \exp \left(K_z - \frac{K_1 + K_2}{4} \right) \cosh \left(K_x - K_y \right) \nonumber \\
       \! \! \!&+& \! \! \! 2 \exp \left(- K_z + \frac{K_1 + K_2}{4} \right) 
       \cosh \left(K_x + K_y \right)\!, \label{bw3} \\
\omega_5 \! \! &=& \! \! \omega_6 =
2 \exp \left(K_z - \frac{K_1 - K_2}{4} \right) \cosh \left(K_x - K_y \right) \nonumber \\
       \! \! \!&+& \! \! \! 2 \exp \left(-K_z + \frac{K_1 - K_2}{4} \right)   
       \cosh \left(K_x + K_y \right)\!, \label{bw5} \\
\omega_7 \! \! &=& \! \! \omega_8 = 
2 \exp \left(K_z + \frac{K_1 - K_2}{4} \right) \cosh \left(K_x - K_y \right) \nonumber \\
       \! \! \!&+& \! \! \! 2 \exp \left(- K_z - \frac{K_1 - K_2}{4} \right)   
       \cosh \left(K_x + K_y \right)\!. \label{bw7}
\end{eqnarray}   
It is quite obvious from the set of Eqs.~(\ref{bw1})-(\ref{bw7}) that the Boltzmann's weights are pairwise equal to each other and there are merely four independent Boltzmann's weights. The spin-1/2 Ising-Heisenberg model defined through the Hamiltonian (\ref{ham}) has been accordingly 
mapped to the eight-vertex model, which quite generally satisfies the Baxter's zero-field condition 
$\omega_1 = \omega_2$, $\omega_3 = \omega_4$, $\omega_5 = \omega_6$, and $\omega_7 = \omega_8$ \cite{baxt82,baxt71,baxt72}. Hence, it follows that the exact solution of the spin-1/2 Ising-Heisenberg model with the pair XYZ Heisenberg interaction and the quartic Ising interactions 
can be extracted from the Baxter's exact solution of the corresponding zero-field eight-vertex model \cite{baxt82,baxt71,baxt72}. For instance, the critical condition of the zero-field eight-vertex model 
\begin{eqnarray}
\omega_1 + \omega_3 + \omega_5 + \omega_7 = 2 \mbox{max} \{\omega_1, \omega_3, \omega_5, \omega_7 \},
\label{cc}
\end{eqnarray}   
directly determines phase transitions of the spin-1/2 Ising-Heisenberg model if the effective Boltzmann's weights (\ref{bw1})-(\ref{bw7}) are substituted into this critical condition. 
It is also worthy to mention that the critical exponents, which characterize the phase transitions 
of the zero-field eight-vertex model, continuously change with the parameter 
$\mu = 2 \arctan(\omega_5 \omega_7/ \omega_1 \omega_3)^{1/2}$ by following the formulas
\begin{eqnarray}
\alpha \! \! \! &=& \! \! \! \alpha' = 2 - \frac{\pi}{\mu}, \quad \beta = \frac{\pi}{16 \mu}, 
\quad \nu = \nu' = \frac{\pi}{2 \mu}, \nonumber \\
\gamma \! \! \! &=& \! \! \! \gamma' = \frac{7 \pi}{8 \mu}, \hspace{0.8cm} \delta = 15, \hspace{0.7cm} \eta = \frac{1}{4}.
\label{ce}
\end{eqnarray} 
Note that the set of relations (\ref{ce}) will also govern changes of the critical exponents 
of the Ising-Heisenberg model provided that the effective Boltzmann's weights (\ref{bw1})-(\ref{bw7}) are used for a calculation of the parameter $\mu$. 

To provide an alternative proof of the exact mapping equivalence between the zero-field eight-vertex model and the spin-1/2 Ising-Heisenberg model, one can utilize the fact that the eight-vertex 
model on a square lattice can also be reformulated as two spin-1/2 Ising square lattices  
coupled together by means of the quartic interaction \cite{wu71,kada71}. In accordance with 
this statement, the effective Boltzmann's factor (\ref{bf}) could be eventually replaced via appropriate algebraic transformation of the form
\begin{eqnarray}
\omega_p \! \! \! \! \! && \! \! \! \! \!
 \left(\sigma^z_{p1}, \sigma^z_{p2}, \sigma^z_{p3}, \sigma^z_{p4} \right) = \nonumber \\
\! \! \! &2& \! \! \!   \exp \left(K_z + K_1 \sigma^z_{p1} \sigma^z_{p3} 
                           + K_2 \sigma^z_{p2} \sigma^z_{p4} \right) \cosh \left(K_x - K_y \right) +
\nonumber \\                           
 \! \! \! &2& \! \! \! \exp \left(- K_z - K_1 \sigma^z_{p1} \sigma^z_{p3} 
                        - K_2 \sigma^z_{p2} \sigma^z_{p4} \right) \cosh \left(K_x + K_y \right)
\nonumber \\ \! \! \! &=& \! \! \!
 R_0 \exp(R_1 \sigma^z_{p1} \sigma^z_{p3} + R_2 \sigma^z_{p2} \sigma^z_{p4} 
       + R_4 \sigma^z_{p1} \sigma^z_{p2} \sigma^z_{p3} \sigma^z_{p4}),
\label{amt} 
\end{eqnarray}
where the mapping parameters $R_1$ and $R_2$ denote the effective pair interactions in two different Ising square lattices and the mapping parameter $R_4$ determines the effective quartic interaction 
that couples together both Ising square lattices. The algebraic transformation (\ref{amt}) must satisfy the 'self-consistency' condition, which means that it must hold independently of spin states of four Ising spins involved therein. The 'self-consistency' condition thus provides a simple connection between the effective Boltzmann's weights (\ref{bw1})-(\ref{bw7}) of the Ising-Heisenberg model 
and the coupling parameters $R_1$, $R_2$, and $R_4$ of the zero-field eight-vertex model 
in the Ising representation
\begin{eqnarray}
R_0 \! \! \! &=& \! \! \! \left(\omega_1 \omega_3 \omega_5 \omega_7 \right)^{1/4}, \label{ev1}  \\
R_1 \! \! \! &=& \! \! \! \ln \left( \frac{\omega_1 \omega_7}{\omega_3 \omega_5} \right), \label{ev2} \\
R_2 \! \! \! &=& \! \! \! \ln \left( \frac{\omega_1 \omega_5}{\omega_3 \omega_7} \right), \label{ev3} \\
R_4 \! \! \! &=& \! \! \! 4 \ln \left( \frac{\omega_1 \omega_3}{\omega_5 \omega_7} \right).
\label{ev4}
\end{eqnarray} 
This is actually an alternative proof of the exact mapping equivalence between the spin-1/2 Ising-Heisenberg model and the zero-field eight-vertex model on a square lattice. 

The usefulness of the latter equivalence consists in that another three useful observations 
can be made from it. First, it is quite evident from Eqs.~(\ref{ev2}) and (\ref{ev3}) that 
the effective pair interactions of both Ising square lattices become equal to each other
($R_1 = R_2$) by imposing the condition $\omega_5 = \omega_7$, which is equivalent 
to $K_1 = K_2$ (or $J_1 = J_2$). This means that a difference in two diagonal quartic 
Ising interactions merely causes a difference in the effective pair interactions of 
the Ising square lattices. Second, the quartic interaction $R_4$ that couples together 
two Ising square lattices vanishes only if at least one from both quartic Ising interactions 
$K_1$ or $K_2$ ($J_1$ or $J_2$) equals zero provided that all interactions are finite. 
Another particular cases with the zero effective quartic interaction $R_4$ are 
the Ising and XY limit of the Heisenberg pair interaction. One actually obtains 
$R_4 \to 0$, $R_1 \to \pm K_1$, and $R_2 \to \pm K_2$ in the Ising limit $K_z \to \pm \infty$, 
whereas $R_4 \to 0$, $R_1 \to -K_1$, and $R_2 \to -K_2$ is obtained in the XY limit 
$K_x, K_y \to \pm \infty$. 
Apart from these rather trivial cases from the Ising universality class, one may expect 
that the spin-1/2 Ising-Heisenberg model will generally exhibit continuously varying critical 
exponents satisfying the weak universality hypothesis \cite{suzu74} due to the non-zero 
effective quartic interaction $R_4$. Third, the spin-1/2 Ising-Heisenberg model defined through 
the Hamiltonian (\ref{ham}) can in turn be refined by the terms depending on the Ising spin
pairs $\sigma^z_{p1} \sigma^z_{p3}$ and $\sigma^z_{p2} \sigma^z_{p4}$ without disturbing 
the exact mapping equivalence to the zero-field eight-vertex model. As a matter of fact, 
an inclusion of two diagonal pair interactions $-J'_1 \sigma^z_{p1} \sigma^z_{p3}$,  
$-J'_2 \sigma^z_{p2} \sigma^z_{p4}$, and the quartic interaction $-J'_4 \sigma^z_{p1} \sigma^z_{p2} \sigma^z_{p3} \sigma^z_{p4}$ into the Hamiltonian (\ref{ham}) merely adds the respective 
coupling constant as an auxiliary constant factor into a definition of the mapping 
parameter (\ref{ev2}), (\ref{ev3}), and (\ref{ev4}), respectively. It is noteworthy, 
however, that the model system is mapped to the more general and yet unsolved eight-vertex model 
by introducing the nearest-neighbor pair interactions $-J''_1 (\sigma^z_{p1} \sigma^z_{p2} + \sigma^z_{p3} \sigma^z_{p4})$ and $-J''_2 (\sigma^z_{p2} \sigma^z_{p3} + \sigma^z_{p1} \sigma^z_{p4})$ 
into the Hamiltonian (\ref{ham}). Even though there does not exist the general exact solution of the corresponding eight-vertex model, it is now well established that the critical exponents 
still vary continuously with the interaction parameters even for this more complex but surely more realistic model \cite{leeu75,nigh77,krin77,swen79,bind80,oitm81,mina93a,mina93b,mina93c}.

\section{Results and discussion}
\label{sec3}

Before proceeding to a discussion of the most interesting results, it is worthy to notice 
that our further analysis will be restricted just to a particular example of the spin-1/2 
Ising-Heisenberg model with the identical quartic Ising interactions 
$J_1 = J_2 = J_4$ and the more symmetric XXZ Heisenberg interaction with $J_x = J_y = J \Delta$, 
$J_z = J$ in order to avoid over-parametrization of the model under investigation. Furthermore, 
it also convenient to normalize all interaction parameters with respect to the $z$-component 
of the XXZ Heisenberg interaction, which will henceforth serve as the energy unit. 
Accordingly, the dimensionless temperature will be set to $k_{\rm B}T/J$, a relative strength of 
the quartic Ising interactions will be proportional to the ratio $J_4/J$, and finally, the parameter 
$\Delta = J_x/J = J_y/J$ will measure a relative strength of the exchange anisotropy 
in the XXZ Heisenberg interaction. 

Let us perform first a comprehensive analysis of the ground state. It is worthwhile to remark that the ground-state spin arrangement will be thoroughly determined by the lowest-energy eigenstate that enters into the greatest Boltzmann's weight, since each Boltzmann's weight involves four eigenenergies that correspond to possible eigenstates 
of the Heisenberg spin pair at a given configuration of the Ising spins (remember that the Ising 
spin configurations are unambiguously assigned to the Boltzmann's weights according to a scheme 
depicted in Fig.~\ref{fig2}). It might be therefore quite useful to quote initially explicit expressions for the effective Boltzmann's weights of simplified version of 
the spin-1/2 Ising-Heisenberg model 
\begin{eqnarray}
\omega_1 \! \! &=& \! \! 2 \exp \! \left[ \frac{\beta \left(2J + J_4 \right)}{8} \right] 
        + 2 \exp \! \left[- \frac{\beta \left(2J + J_4 \right)}{8}  \right] \nonumber \\
        \! \! &\times& \! \! \cosh \left(\frac{\beta J \Delta }{2} \right), \nonumber \\
\omega_3 \! \! &=& \! \! 2 \exp \! \left[ \frac{\beta \left(2J - J_4 \right)}{8} \right] 
        + 2 \exp \! \left[- \frac{\beta \left(2J - J_4 \right)}{8}  \right] \nonumber \\
        \! \! &\times& \! \! \cosh \left(\frac{\beta J \Delta }{2} \right), \label{bwxxz} \\
\omega_5 \! \! &=& \! \! \omega_7 = 2 \exp \! \left(\frac{\beta J}{4} \right) \! + 2 \exp \! 
        \left(- \frac{\beta J}{4} \right) \! \cosh \! \left(\frac{\beta J \Delta}{2} \right). \nonumber
\end{eqnarray} 
It directly follows from the set of Eqs.~(\ref{bwxxz}) that the greatest Boltzmann's weight is either $\omega_1$ or $\omega_3$. Besides, another useful observation is that both these Boltzmann's weights are simply connected through the relation $\omega_1 (\pm J_4) = \omega_3 (\mp J_4)$ and one may 
further assume the positive quartic Ising interaction $J_4>0$ without loss of generality. This simplification is a direct consequence of the affirmation that the investigated model system 
is invariant under the transformation $J_4 \to -J_4$ and $(\sigma_{p1}^z, \sigma_{p2}^z, 
\sigma_{p3}^z, \sigma_{p4}^z) \to (\sigma_{p1}^z, \sigma_{p2}^z, -\sigma_{p3}^z, -\sigma_{p4}^z)$.
The ground-state phase diagram must be therefore symmetric with respect to the $J_4 = 0$ axis 
and the respective spin arrangements to emerge in a sector $J_4<0$ can be obtained from 
its symmetry related counterpart in a sector $J_4>0$ by a mere interchange of the Ising spin configurations inherent to the Boltzmann's weights $\omega_1$ and $\omega_3$. 

For better orientation, we will distinguish our subsequent analysis of the spin-1/2 Ising-Heisenberg model with the ferromagnetic ($J>0$) and antiferromagnetic ($J<0$) Heisenberg pair interaction, respectively. In the former case, the greatest Boltzmann's weight is $\omega_1$ for $\Delta<1$ and $\omega_3$ for $\Delta>1$, under the assumptions $T=0$ and $J_4>0$ (note that the reverse conditions hold for the case $J_4<0$). The global ground-state phase
\begin{figure}[htb]
\vspace{-0.3cm}
\begin{center}
\includegraphics[width=8cm]{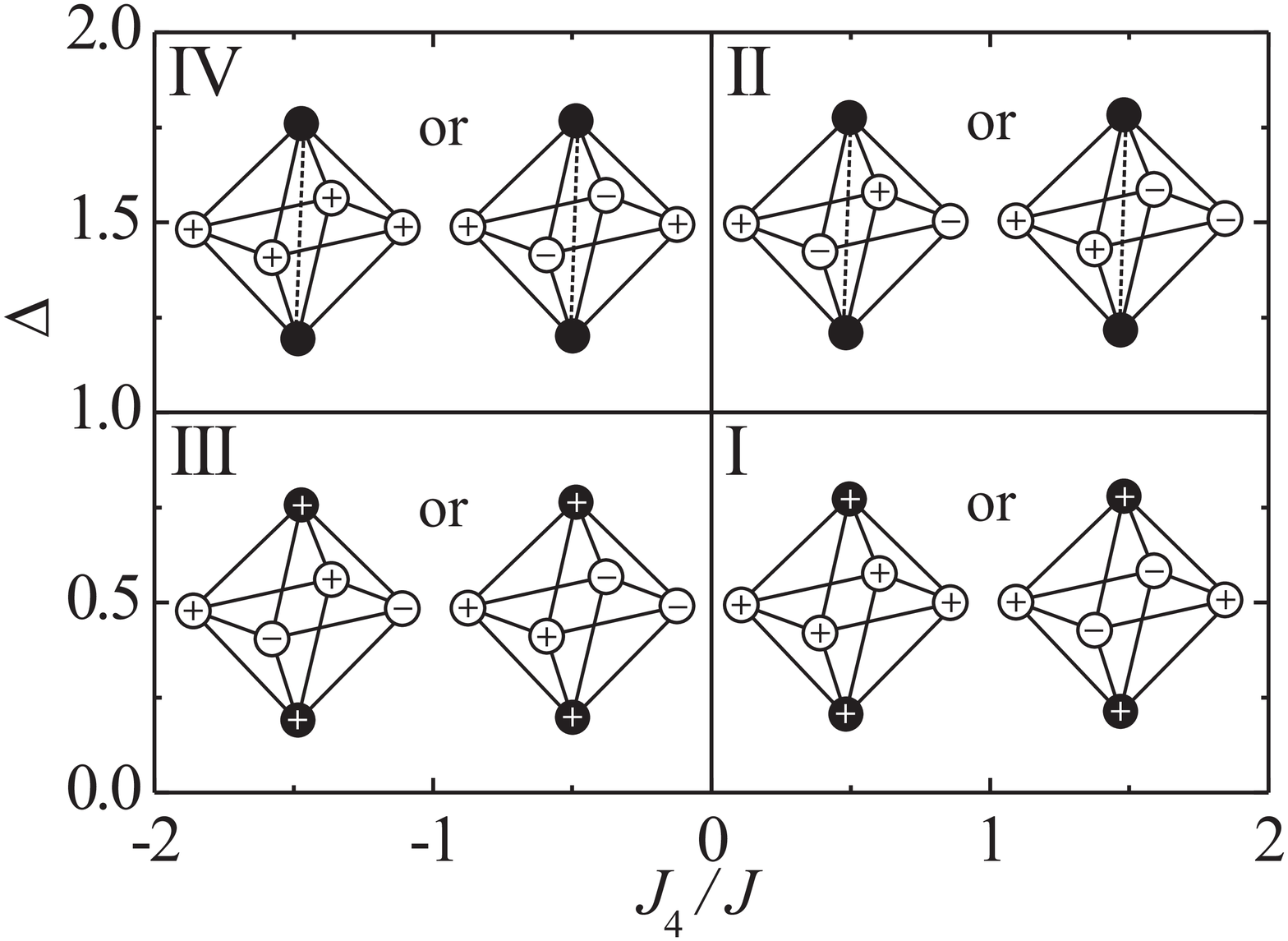}
\end{center}
\vspace{-1.2cm}
\caption{Ground-state phase diagram of the spin-1/2 Ising-Heisenberg model with the ferromagnetic interaction $J>0$ in the $J_4/J-\Delta$ plane. Each sector of the phase diagram contains typical 
spin configurations of the Ising and Heisenberg spins, which form an elementary octahedron. 
The sign '$\pm$' denotes spin states $\sigma^z = \pm 1/2$ and $S^z = \pm 1/2$ of the Ising and Heisenberg spins, respectively. Broken lines, which connect both apical Heisenberg spins in 
the phases $|{\rm II} \rangle$ and $|{\rm IV} \rangle$, label the entangled spin state 
$\left(|+,- \rangle + |-,+ \rangle \right)/\sqrt{2}$.}
\label{fig3}
\end{figure}
diagram of the spin-1/2 Ising-Heisenberg model with the ferromagnetic Heisenberg interaction is displayed in Fig.~\ref{fig3}. Clearly, the ground-state diagram constitute four different phases 
\begin{eqnarray}
|{\rm I} \rangle \! \! \! &=& \! \! \! \prod_p |+,\pm,+,\pm \rangle_{\sigma_p} |+,+ \rangle_{S_p}, 
\label{gsa} \\
|{\rm II} \rangle \! \! \! &=& \! \! \! \prod_p |+,\pm,-,\mp \rangle_{\sigma_p} 
\frac{1}{\sqrt{2}} \left(|+,- \rangle + |-,+ \rangle \right)_{S_p}\!, 
\label{gsb} \\
|{\rm III} \rangle \! \! \! &=& \! \! \! \prod_p |+,\pm,-,\mp \rangle_{\sigma_p} |+,+ \rangle_{S_p},  \label{gsc} \\
|{\rm IV} \rangle \! \! \! &=& \! \! \! \prod_p |+,\pm,+,\pm \rangle_{\sigma_p} 
\frac{1}{\sqrt{2}} \left(|+,- \rangle + |-,+ \rangle \right)_{S_p}\!, 
\label{gsd}
\end{eqnarray} 
which are written in a form of the product over the respective lowest-energy eigenstate of the Hamiltonian (\ref{ham}) that is of course identical for each elementary unit cell (octahedron). 
Note furthermore that the first (second) ket vector behind the product symbol determines spin 
states of the Ising (Heisenberg) spins and another equivalent representations of the eigenstates 
can be obtained from the eigenvectors (\ref{gsa})-(\ref{gsd}) under the reversal of all Ising 
and/or Heisenberg spins. The phases $|{\rm I} \rangle$ and $|{\rm III} \rangle$ are consequently eight-fold degenerate as spin reversals of the Ising and Heisenberg spins can be performed independently of each other, whereas the phases $|{\rm II} \rangle$ and $|{\rm IV} \rangle$ 
are just four-fold degenerate as the spin reversal of the Heisenberg spins does not 
in fact lead to a new eigenstate. 

It is worthwhile to remark that another two obvious features directly follow from Fig.~\ref{fig3} 
and Eqs.~(\ref{gsa})-(\ref{gsd}). The Heisenberg spins are ferromagnetically aligned with respect 
to each other in the phases $|{\rm I} \rangle$ and $|{\rm III} \rangle$ where $\Delta<1$, 
while they reside the entangled spin state $\left(|+,- \rangle + |-,+ \rangle \right)/\sqrt{2}$ 
in the phases $|{\rm II} \rangle$ and $|{\rm IV} \rangle$ where $\Delta>1$. In addition, 
there appears a perfect ferromagnetic or antiferromagnetic alignment on a square lattice 
of the Ising spins in the phases $|{\rm I} \rangle$ and $|{\rm IV} \rangle$ in contrast to 
two feasible superantiferromagnetic orderings that occur in the phases $|{\rm II} \rangle$ 
and $|{\rm III} \rangle$, where a perfect ferromagnetic arrangement of the Ising spins 
in a horizontal direction is accompanied with a perfect antiferromagnetic arrangement 
in a vertical direction, or vice versa. The overall understanding of a nature of 
the aforedescribed spin arrangements readily follows from the energy spectrum of the spin-1/2 
XXZ Heisenberg dimer. Namely, the lowest-energy eigenstate of the spin-1/2 XXZ Heisenberg dimer 
is the ferromagnetic state $|+,+ \rangle$ if one considers the easy-axis exchange anisotropies $\Delta<1$, while the entangled spin state $\left(|+,- \rangle + |-,+ \rangle \right)/\sqrt{2}$ 
becomes the lowest-energy eigenstate for the easy-plane exchange anisotropies $\Delta>1$. 
The relevant spin arrangement of the Ising spins is subsequently driven by the effort 
to minimize the energy gain arising from the quartic Ising interactions. Thus, there must be 
either zero or two spins with opposite orientation with respect to the others among the four 
spins (two Ising and two Heisenberg) involved in the positive quartic Ising interaction $J_4>0$, 
whereas there must be just one unaligned spin among them whenever the negative quartic Ising interaction $J_4<0$ is assumed. 

Surprisingly, the ground-state analysis becomes much more simple for the spin-1/2 Ising-Heisenberg model with the antiferromagnetic ($J<0$) Heisenberg interaction. It is quite apparent from the set 
of Eqs.~(\ref{bwxxz}) that the greatest Boltzmann's weight is always being $\omega_3$ ($\omega_1$) 
if one considers the positive (negative) quartic Ising interaction. Owing to this fact, 
the overall ground-state phase diagram constitute just two different phases  
\begin{eqnarray}
|{\rm V} \rangle \! \! \! &=& \! \! \! \prod_p |+,\pm,-,\mp \rangle_{\sigma_p} 
\frac{1}{\sqrt{2}} \left(|+,- \rangle - |-,+ \rangle \right)_{S_p}\!, 
\label{gse} \\
|{\rm VI} \rangle \! \! \! &=& \! \! \! \prod_p |+,\pm,+,\pm \rangle_{\sigma_p} 
\frac{1}{\sqrt{2}} \left(|+,- \rangle - |-,+ \rangle \right)_{S_p}\!, 
\label{gsf}
\end{eqnarray} 
whereas the phase $|{\rm V} \rangle$ is the ground state for $J_4>0$ and the phase $|{\rm VI} \rangle$ for $J_4<0$. It should be stressed that both four-fold degenerate phases $|{\rm V} \rangle$ and $|{\rm VI} \rangle$ quite closely resemble the phases $|{\rm II} \rangle$ and $|{\rm IV} \rangle$ described previously by the analysis of the ferromagnetic model. As a matter of fact, the only difference 
between the phases $|{\rm V} \rangle$ and $|{\rm II} \rangle$ (or $|{\rm VI} \rangle$ and $|{\rm IV} \rangle$) consists in a quantum entanglement of the Heisenberg spin pairs, which is described 
by the antisymmetric singlet-dimer wave function $\left(|+,- \rangle - |-,+ \rangle \right)/\sqrt{2}$ in the antiferromagnetic model with $J<0$ and its symmetric counterpart $\left(|+,- \rangle + |-,+ \rangle \right)/\sqrt{2}$ in the ferromagnetic model with $J>0$.

Now, let us turn our attention to a detailed analysis of the finite-temperature phase diagrams. 
It is worthwhile to recall that phase transition lines of our simplified version of the spin-1/2 Ising-Heisenberg model can be straightforwardly obtained from the critical condition of the corresponding zero-field eight-vertex model by substituting the effective Boltzmann's weights (\ref{bwxxz}) into Eq.~(\ref{cc}). It should be also mentioned that the greatest Boltzmann's weight might be either $\omega_1$ or $\omega_3$. In the former case ($\omega_1 > \omega_3$), the critical condition reads
\begin{eqnarray}
\cosh \left(\frac{\beta_{\rm c} J}{2} \Delta \right) = \exp \left(\frac{\beta_{\rm c} J}{2} \right) 
\frac{\sinh \left(\frac{\beta_{\rm c} J_4}{8} \right) - 1}
     {\sinh \left(\frac{\beta_{\rm c} J_4}{8} \right) + 1},
\label{cclw}
\end{eqnarray} 
while in the latter case ($\omega_1 < \omega_3$) the critical condition becomes
\begin{eqnarray}
\cosh \left(\frac{\beta_{\rm c} J}{2} \Delta \right) = \exp \left(\frac{\beta_{\rm c} J}{2} \right) 
\frac{\sinh \left(\frac{\beta_{\rm c} J_4}{8} \right) + 1}
     {\sinh \left(\frac{\beta_{\rm c} J_4}{8} \right) - 1},
\label{ccrw}
\end{eqnarray}
where $\beta_{\rm c} = 1/(k_{\rm B} T_{\rm c})$ and $T_{\rm c}$ is the critical temperature.
Interestingly, both critical conditions (\ref{cclw}) and (\ref{ccrw}) can be joined together 
to yield a single critical condition 
\begin{eqnarray}
\sinh \left(\frac{\beta_{\rm c} |J_4|}{8} \right) 
   = \frac{\exp \left(\frac{\beta_{\rm c} J}{2} \right) 
   + \cosh \left(\frac{\beta_{\rm c} J}{2} \Delta \right)}
   {\left| \exp \left( \frac{\beta_{\rm c} J}{2} \right)
   - \cosh \left(\frac{\beta_{\rm c} J}{2} \Delta \right) \right|},
\label{ccgen}
\end{eqnarray}
which is valid in a whole region of the parameter space. The absolute value of the quartic 
Ising interaction appears in Eq.~(\ref{ccgen}) as a direct consequence of the symmetry relation 
between the Boltzmann's weights $\omega_1 (\pm J_4) = \omega_3 (\mp J_4)$, which are mutually interchangeable under the transformation $J_4 \to -J_4$. Accordingly, the critical temperature 
of the spin-1/2 Ising-Heisenberg model must be independent of the sign of the quartic Ising 
interaction and one may still consider positive values of the quartic Ising interaction $J_4>0$ without 
loss of the generality. Henceforth, we will therefore restrict ourselves just to an analysis of 
the critical lines of the phases $|{\rm I} \rangle$, $|{\rm II} \rangle$, and $|{\rm V} \rangle$ 
emerging for the positive quartic Ising interaction $J_4>0$. It should be nevertheless mentioned
that the critical lines for the phases $|{\rm III} \rangle$, $|{\rm IV} \rangle$, and $|{\rm VI} \rangle$ appearing in the case $J_4<0$ are formally identical with the displayed critical lines 
for the phases $|{\rm I} \rangle$, $|{\rm II} \rangle$, and $|{\rm V} \rangle$, respectively.

\begin{figure}[htb]
\begin{center}
\includegraphics[width=7.5cm]{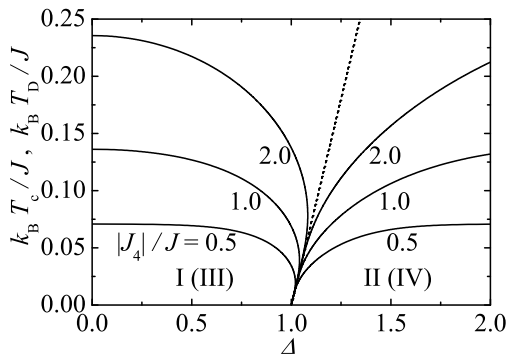}
\end{center}
\vspace{-1.1cm}
\caption{Solid lines depict the critical temperature as a function of the exchange anisotropy 
for the spin-1/2 Ising-Heisenberg model with the ferromagnetic Heisenberg interaction $J>0$ 
at three different relative strengths of the quartic Ising interaction. The left wing of critical 
lines determines phase boundaries for the phases $|{\rm I} \rangle$ ($J_4>0$) or $|{\rm III} \rangle$ ($J_4<0$), while the right wing determines phase boundaries for the phases $|{\rm II} \rangle$ 
($J_4>0$) or $|{\rm IV} \rangle$ ($J_4<0$). The broken line shows the disorder solution obtained by solving Eq.~(\ref{ds}).}
\label{fig4}
\end{figure}

Let us construct first the finite-temperature phase diagram of the spin-1/2 Ising-Heisenberg model 
with the ferromagnetic Heisenberg interaction. The greatest Boltzmann's weight for the ferromagnetic model with $J>0$ is $\omega_1$ if and only if
\begin{eqnarray}
\mbox{sgn}\left(J_4 \right) \left[\exp \left(\frac{\beta J}{2} \right) 
- \cosh \left(\frac{\beta J}{2} \Delta \right) \right]  > 0, 
\label{nsc}
\end{eqnarray}
otherwise $\omega_3$ becomes the greatest Boltzmann's weight. Consequently, the critical condition (\ref{cclw}) determines phase transitions just if the inequality (\ref{nsc}) holds, while the 
critical condition (\ref{ccrw}) is valid in the rest of the parameter space. Phase transition 
lines in the form of critical temperature versus exchange anisotropy dependences are drawn 
in Fig.~\ref{fig4} for three different relative strengths of the quartic Ising interaction. 
As one can see, the critical temperature generally exhibits a remarkable dependence on the exchange anisotropy with two marked wings of critical lines that merge together at the ground-state boundary 
$\Delta=1$ between the phases $|{\rm I} \rangle$ and $|{\rm II} \rangle$.
It should be mentioned that the left wing of displayed phase boundaries is a solution of 
the critical condition (\ref{cclw}), while the right wing is a solution of the critical condition (\ref{ccrw}). In this regard, the left wings deliminate critical lines of the phase $|{\rm I} \rangle$, whereas the right wings represent critical lines of the phase $|{\rm II} \rangle$. 
The critical temperature of the phase $|{\rm I} \rangle$ at first steadily decreases as 
the exchange anisotropy $\Delta$ is raised from zero, then it exhibits an interesting re-entrant behavior before it finally tends to zero in the limit $\Delta \to 1$. It is noticeable that the re-entrant phase transitions are observable just in a relatively narrow interval of the exchange anisotropies $\Delta \in (1, \Delta_{\rm max})$, whose upper bound $\Delta_{\rm max}$ is 
the greater, the greater a relative strength of the quartic Ising interaction is. Contrary to this, 
the right wing of critical lines, which allocate phase transitions of the phase $|{\rm II} \rangle$, characterizes a monotonous increase of the critical temperature with the exchange anisotropy. 
An origin of the observed re-entrant transitions obviously lies in two times higher degeneracy 
of the phase $|{\rm I} \rangle$ compared to the phase $|{\rm II} \rangle$. Accordingly, 
two successive re-entrant transitions from the paramagnetic phase to the phase $|{\rm I} \rangle$ 
and back may take place in the parameter region, where the phase $|{\rm II} \rangle$ constitutes the ground state and the phase $|{\rm I} \rangle$ has slightly higher energy, on account of the much larger entropy gain of the phase $|{\rm I} \rangle$ obtained upon the temperature increase. It is worthy to mention that similar re-entrant phase transitions have been found in a variety of frustrated Ising models, which have been exactly solved by establishing a precise mapping relationship with the corresponding free-fermion vertex models \cite{diep04,vaks66,chik87,azar87,azar89,diep91,deba91}.
To bring a deeper insight into the re-entrant phenomenon, it is also worthwhile to inspect 
the disorder solution to be derived from the condition    
\begin{eqnarray}
\omega_1 = \omega_3 \Leftrightarrow 
\exp \left( \frac{\beta_{\rm D} J}{2} \right) = \cosh \left(\frac{\beta_{\rm D} J \Delta}{2} \right), 
\label{ds}
\end{eqnarray}
where $\beta_{\rm D} = 1/(k_{\rm B} T_{\rm D})$ and $T_{\rm D}$ is the disorder temperature. 
The disorder solution entails an effective reduction of the dimensionality and ensures 
a disordered nature of the spin system on the particular subvariety of the parameter space 
given by Eq.~(\ref{ds}) \cite{diep04,azar87,azar89,diep91,deba91,step70a,step70b,step70c}. 
It is quite apparent from Fig.~\ref{fig4} that the disorder (broken) line calculated 
from the condition (\ref{ds}) has a finite positive tangent, which can be approximated 
in a relatively wide temperature range by a linear dependence of the disorder temperature 
$T_{\rm D}$ on the exchange anisotropy $\Delta$ through the relation 
$k_{\rm B} T_{\rm D}/J = (\Delta - 1)/\ln 4$.

\begin{figure}[htb]
\begin{center}
\includegraphics[width=7cm]{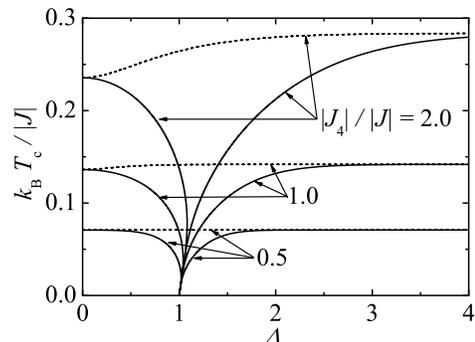}
\end{center}
\vspace{-1.2cm}
\caption{The critical temperature as a function of the exchange anisotropy for the spin-1/2 Ising-Heisenberg model with the ferromagnetic (solid lines) or antiferromagnetic (broken lines) Heisenberg pair interaction at three different relative strengths of the quartic Ising interaction.}
\label{fig5}
\end{figure}

For the sake of comparison, the critical temperatures of the spin-1/2 Ising-Heisenberg model 
with both ferromagnetic ($J>0$) as well as antiferromagnetic ($J<0$) Heisenberg pair interaction 
are depicted in Fig.~\ref{fig5} for three different relative strengths of the quartic Ising interaction. The critical lines of the ferromagnetic model are displayed as solid lines,
while the critical lines of the antiferromagnetic model are drawn as broken lines. 
The displayed critical lines of the antiferromagnetic model in fact determine critical temperatures 
of the phase $|{\rm V} \rangle$, which is the only possible ground state when $J<0$ and $J_4>0$. Evidently, the antiferromagnetic model generally exhibits a rather small variation 
of the critical temperature upon varying the exchange anisotropy in comparison with the marked 
two-wing dependence of the critical temperature that shows the ferromagnetic model. Another 
interesting fact to observe here, and also derived from Eq.~(\ref{ccgen}), is that the critical temperatures of the ferromagnetic and antiferromagnetic model become equal to each other 
in two limiting cases $\Delta \to 0$ and $\Delta \to \infty$. 
From this perspective, the most obvious difference between the critical temperatures of the 
ferromagnetic and antiferromagnetic Ising-Heisenberg models appears in a vicinity of the isotropic 
limit $\Delta=1$, where the critical temperature of the ferromagnetic model asymptotically 
vanishes due to the zero-temperature phase transition between the phases $|{\rm I} \rangle$ 
and $|{\rm II} \rangle$.

\begin{figure}[htb]
\hspace{-0.7cm}
\begin{center}
\includegraphics[width=9cm]{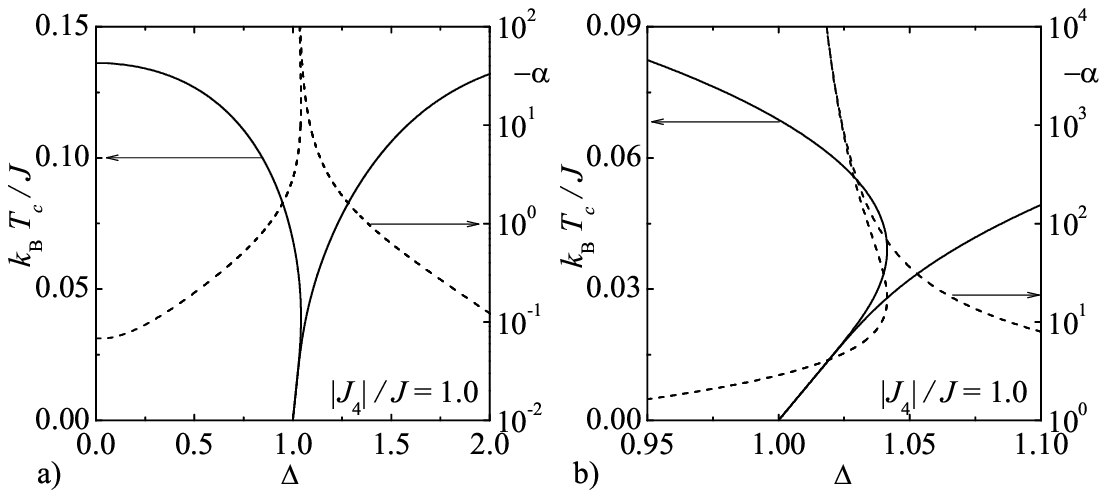}
\end{center}
\vspace{-1.0cm}
\caption{The changes of the critical exponent $\alpha$ along the critical line of the spin-1/2 Ising-Heisenberg model with the ferromagnetic interaction $J>0$ and $|J_4|/J = 1.0$. Solid lines, 
which are scaled with respect to left axes, show the critical temperature as a function of 
the exchange anisotropy. Broken lines, which are scaled with respect to right axes, 
display the relevant changes of the critical exponent $-\alpha$ in a semi-logarithmic scale. 
Fig.~\ref{fig6}b) illustrates a detail of the parameter region, where both re-entrant phase 
transitions as well as a quantum critical point occur.}
\label{fig6}
\end{figure}

At this point, let us make certain comments on changes of critical exponents along the critical 
lines. For this purpose, typical changes of the critical exponent $\alpha$ along the phase 
transition line of the spin-1/2 Ising-Heisenberg model with the ferromagnetic Heisenberg interaction
and $|J_4|/J = 1$ are depicted in Fig.~\ref{fig6}. Solid lines display a variation of the critical temperature with the exchange anisotropy, which are scaled with respect to left axes, while broken lines scaled with respect to right axes show the relevant changes of the critical exponent $\alpha$. 
As could be expected, the critical exponent $\alpha$ varies continuously along the line of critical points according to the relations (\ref{ce}), which connect changes of the critical exponents to respective changes of the interaction parameters through the parameter $\mu$. It turns out that 
the investigated model system generally exhibits rather smooth continuous phase transitions, 
since the order of phase transition is proportional to $r=2-\alpha$ (see for instance pp.~16--17 
in Ref. \onlinecite{baxt82}) and the critical exponent $\alpha$ lies within the range 
$\alpha \in (-\infty, 0)$. 
The most striking finding is, however, that the critical exponent $\alpha$ exhibits a peculiar singularity in a vicinity of the ground-state boundary between the phases $|{\rm I} \rangle$ 
and $|{\rm II} \rangle$ where $\alpha \to - \infty$. Owing to this fact, the zero-temperature phase transition between the phases $|{\rm I} \rangle$ and $|{\rm II} \rangle$ emerging at $\Delta=1$ 
is of infinite order and hence, this special critical point actually represents a remarkable 
quantum critical point. It is also worthy to notice that the displayed variations of the critical exponent $- \alpha$ quite closely resemble dependences of other critical exponents 
$\beta$, $\gamma$ and $\nu$ due to a similar mathematical structure of the relations (\ref{ce}). 

\begin{figure}[htb]
\begin{center}
\includegraphics[width=7.5cm]{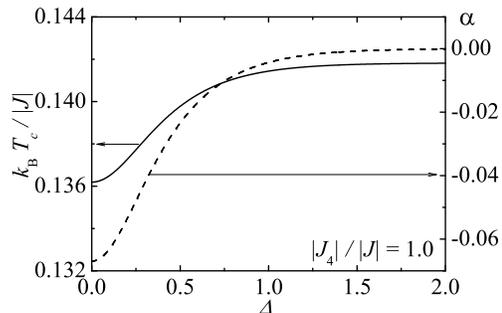}
\end{center}
\vspace{-1.1cm}
\caption{The changes of the critical exponent $\alpha$ along the critical line of the spin-1/2 Ising-Heisenberg model with the antiferromagnetic interaction $J<0$ and $|J_4|/|J| = 1.0$. 
The solid line, which is scaled with respect to the left axis, shows the critical temperature 
as a function of the exchange anisotropy. The broken line, which is scaled with respect 
to the right axis, displays the relevant changes of the critical exponent $\alpha$ 
along this critical line.}
\label{fig7}
\end{figure}

Furthermore, the changes of the critical exponent $\alpha$ along the critical line of the spin-1/2 Ising-Heisenberg model with the antiferromagnetic Heisenberg interaction are shown in Fig.~\ref{fig7}.
In this particular case, the changes of the critical exponent $\alpha$ are restricted to 
a rather narrow finite interval even though a negative value of the critical exponent 
$\alpha$ still implies a continuous nature of the phase transitions $r \geq 2$. 
It is also quite interesting to notice that the critical exponent $\alpha$ monotonically 
increases from its smallest value at $\Delta = 0$ before it gradually tends towards 
$\alpha \to 0$ in the limit $\Delta \to \infty$. Altogether, the critical lines of both 
ferromagnetic as well as antiferromagnetic model turned out to be the lines of continuous 
phase transitions of the order $r \geq 2$, but the ferromagnetic model generally exhibits 
much more pronounced changes of the critical exponents including a rather strange quantum 
critical point of the infinite order with diverging critical exponents.

Before concluding, let us make few remarks on a critical behavior of another particular 
case of the spin-1/2 Ising-Heisenberg model with $J_x = J_y = J \Delta$, $J_z = J$
and the quartic Ising interactions $J_1 = -J_2 = J'_4$. This particular case differs 
from the previous one just in a nature of the quartic Ising interactions, which are 
of equal relative strengths but of different signs. In this respect, the positive quartic 
interaction will prefer spin alignments with either zero or two opposite spins among 
the four spins (two Ising and two Heisenberg) involved therein, while the negative 
quartic interaction will favour spin alignments with just one unaligned spin among them. 
Interestingly, the phase diagrams of this particular case will be essentially identical 
with the ones discussed previously, since there is a simple relation between the effective 
Boltzmann's weights of both particular cases. As a matter of fact, it directly follows 
from Eqs.~(\ref{bw1})-(\ref{bw7}) that the effective Boltzmann's weights of the model 
with two non-uniform quartic interactions $J_1 = -J_2 = J'_4$ can be expressed through 
the Boltzmann's weights (\ref{bwxxz}) of the uniform case as $\omega'_1 = \omega'_3 = \omega_5$, 
$\omega'_5 = \omega_3$, and $\omega'_7 = \omega_1$. It is quite apparent that the role 
of different Boltzmann's weights is merely interchanged and the greatest Boltzmann's weight 
is now being either $\omega'_7$ or $\omega'_5$. However, this fact does not affect the 
ground-state and finite-temperature phase diagrams because the critical condition (\ref{cc}) 
is quite symmetric with respect to all four Boltzmann's weights involved therein. Hence, 
it follows that the zero- and finite-temperature phase diagrams shown in Figs.~\ref{fig3}-\ref{fig5} remain valid and one should only perform a respective change of the Ising spin configurations 
from $\omega_1 \to \omega'_7$ and $\omega_3 \to \omega'_5$, respectively. The ground-state phase diagram of the ferromagnetic ($J>0$) model with 
the non-uniform quartic interactions, which is shown in Fig.~\ref{fig3}, thus constitute the phases
\begin{eqnarray}
|{\rm I'} \rangle \! \! \! &=& \! \! \! \prod_p |+,\mp,+,\pm \rangle_{\sigma_p} |+,+ \rangle_{S_p}, 
\label{gsac} \\
|{\rm II'} \rangle \! \! \! &=& \! \! \! \prod_p |\pm,+,\mp,+ \rangle_{\sigma_p} 
\frac{1}{\sqrt{2}} \left(|+,- \rangle + |-,+ \rangle \right)_{S_p}\!, 
\label{gsbc} \\
|{\rm III'} \rangle \! \! \! &=& \! \! \! \prod_p |\pm,+,\mp,+ \rangle_{\sigma_p} |+,+ \rangle_{S_p},  \label{gscc} \\
|{\rm IV'} \rangle \! \! \! &=& \! \! \! \prod_p |+,\mp,+,\pm \rangle_{\sigma_p} 
\frac{1}{\sqrt{2}} \left(|+,- \rangle + |-,+ \rangle \right)_{S_p}\!. 
\label{gsdc}
\end{eqnarray}     
while the ground state of the antiferromagnetic ($J<0$) model is either being $|{\rm V'} \rangle$ 
for $J'_4>0$ or $|{\rm VI'} \rangle$ for $J'_4<0$
\begin{eqnarray}
|{\rm V'} \rangle \! \! \! &=& \! \! \! \prod_p |\pm,+,\mp,+ \rangle_{\sigma_p} 
\frac{1}{\sqrt{2}} \left(|+,- \rangle - |-,+ \rangle \right)_{S_p}\!, 
\label{gsec} \\
|{\rm VI'} \rangle \! \! \! &=& \! \! \! \prod_p |+,\mp,+,\pm \rangle_{\sigma_p} 
\frac{1}{\sqrt{2}} \left(|+,- \rangle - |-,+ \rangle \right)_{S_p}\!. 
\label{gsfc}
\end{eqnarray} 

\begin{figure}[htb]
\begin{center}
\includegraphics[width=9cm]{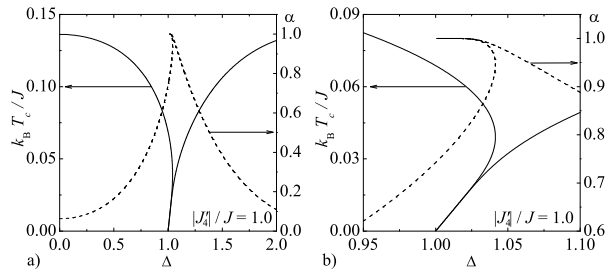}
\end{center}
\vspace{-1.1cm}
\caption{The changes of the critical exponent $\alpha$ along the critical line of the spin-1/2 Ising-Heisenberg model with the ferromagnetic interaction $J>0$ and the relative strength of 
the non-uniform quartic interactions $|J'_4|/J = 1.0$. 
Solid lines, which are scaled with respect to left axes, show the critical temperature 
as a function of the exchange anisotropy. Broken line, which are scaled with respect 
to right axes, display the relevant changes of the critical exponent $\alpha$.}
\label{fig8}
\end{figure}

Despite similarity of the ground-state and finite-temperature phase diagrams, the critical 
exponents of the particular case with the non-uniform quartic interactions 
are fundamentally different from the ones of the former particular case with two identical 
quartic interactions. Namely, the changes of the critical exponents depend through the set 
of Eqs.~(\ref{ce}) on the parameter $\mu$, which is not symmetric with respect to all four 
Boltzmann's weights. Therefore, it might be quite interesting to ascertain how the critical 
exponents change by assuming two quartic interactions of different sign (nature). The variations 
of the critical exponent $\alpha$ along the critical line of the ferromagnetic model ($J>0$)
are displayed in Fig.~\ref{fig8} for the particular case with $|J'_4|/J = 1$.
It is quite obvious from this figure that the critical exponent $\alpha$ exhibits an outstanding dependence with the global maximum $\alpha_{\rm max} = 1$ emerging at $\Delta = 1$. As a result, 
the zero-temperature phase transition between the phases $|{\rm I'} \rangle$ and $|{\rm II'} \rangle$ (or $|{\rm III'} \rangle$ and $|{\rm IV'} \rangle$) is in fact a discontinuous first-order phase transition ($r=1$) unlike the aforedescribed continuous phase transition of the infinite order between 
the phases $|{\rm I} \rangle$ and $|{\rm II} \rangle$ (or $|{\rm III} \rangle$ and $|{\rm IV} \rangle$).
Besides, the critical exponent $\alpha$ evidently acquires positive values from the interval 
$\alpha \in (0,1)$, which indicates discontinuous nature of the phase transitions ($r < 2$) 
along the whole critical line of the ferromagnetic model with the non-uniform quartic interactions. 

\begin{figure}[htb]
\begin{center}
\includegraphics[width=7.5cm]{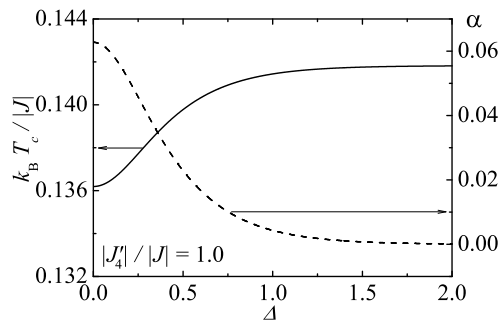}
\end{center}
\vspace{-1.1cm}
\caption{The changes of the critical exponent $\alpha$ along the critical line of the spin-1/2 Ising-Heisenberg model with the antiferromagnetic interaction $J<0$ and the relative strength
of the non-uniform quartic interactions $|J'_4|/|J| = 1.0$. 
The solid line, which is scaled with respect to the left axis, shows the critical temperature 
as a function of the exchange anisotropy. The broken line, which is scaled with respect 
to the right axis, displays the relevant changes of the critical exponent $\alpha$.}
\label{fig9}
\end{figure}

For completeness, we depict in Fig.~\ref{fig9} the critical exponent $\alpha$ of the 
antiferromagnetic model ($J<0$) with the same relative strength of non-uniform quartic 
interactions $|J'_4|/|J| = 1$.
It can be clearly seen from this figure that the critical exponent $\alpha$ monotonically decreases with the increase of the exchange anisotropy even although it always remains positive. It is noteworthy that the precisely opposite trends were observed in the former particular case with the uniform quartic interactions and there are only two common features of both particular cases with the antiferromagnetic pair interaction $J<0$. First, the changes of the critical exponent $\alpha$ are restricted  just 
to a rather narrow finite interval and second, the same asymptotic value $\alpha = 0$ 
is achieved in the limit $\Delta \to \infty$. Altogether, it could be concluded that the Ising-Heisenberg model with the non-uniform quartic interactions generally exhibits 
a discontinuous phase transitions of the order $r<2$ no matter whether the ferromagnetic 
or antiferromagnetic nature of the Heisenberg pair interaction $J$ is assumed. 

\section{Conclusions}
\label{sec4}

In the present work, we have furnished proof of an exact mapping equivalence between the spin-1/2 Ising-Heisenberg model on a two-dimensional lattice of edge-sharing octahedrons and the zero-field eight-vertex model on a square lattice. In accordance with this exact mapping correspondence, 
the critical behavior of the model under investigation closely resembles an outstanding critical behavior of the zero-field eight-vertex model and Ashkin-Teller model \cite{baxt82}, which exhibit 
critical lines with continuously varying critical exponents satisfying the weak universality 
hypothesis \cite{suzu74}. It is worthwhile to remark, moreover, that the similar precise 
mapping relationship between the spin-1/2 Ising-Heisenberg model on a square-hexagon lattice 
and the zero-field eight-vertex model was recently found by Valverde \textit{et al}. 
in a restricted region of the interaction parameters \cite{valv08}. Unlike this case, 
the exact mapping correspondence with the zero-field eight-vertex model holds in our case 
quite generally, i.e. it is not restricted to any particular subvariety of the parameter space.

Exact results for the spin-1/2 Ising-Heisenberg model with the pair XYZ Heisenberg and 
quartic Ising interactions imply that the model with the antiferromagnetic pair interaction 
surprisingly exhibits less significant changes of both critical temperatures as well as 
critical exponents than the model with the ferromagnetic pair interaction. The most interesting 
finding to emerge from the present study is however an exact evidence of a quantum critical point 
of the infinite order, which characterizes a peculiar singular behavior of the critical exponents 
in a close vicinity of the isotropic limit of the Heisenberg pair interaction. In addition, 
it was shown that the critical exponents vary continuously over the entire range of allowed values 
by changing the exchange anisotropy in the Heisenberg pair interaction and the relative 
strength of the quartic interaction. From this point of view, the investigated Ising-Heisenberg 
model represents a rare example of the exactly solved quantum spin model with such an unusual 
weak-universal critical behavior. 

Next, it is worth noticing that the Ising-Heisenberg model with the pair Heisenberg 
and quadratic Ising interactions can also be interpreted as an interesting example of 
the exactly solved quantum dimer system with a rather complicated even-body 
interactions between dimers. This means that the system of quantum dimers in some specific 
fluctuations or fields is also equivalent with the exactly solved zero-field eight-vertex model, 
which contradicts the standard universality hypothesis in that its critical exponents 
vary continuously with the interaction parameters. In addition, the quartic and other 
even-body interactions turned out to play an important role in determining magnetic 
properties of several insulating magnetic materials \cite{roge89,suga90,hond93,mizu99,cold01,kata02,mats00,notb07}, which makes 
the presented exact results more interesting also from the experimental point of view.   
Even though it would be rather striking coincidence if some real magnetic material 
would obey very specific topological requirements of the model under investigation, 
it is quite reasonable to suspect that our exact results might at least shed light 
on some important aspects of the critical behavior of real magnetic materials. 

Last but not least, it should be mentioned that it could be quite interesting to explore also temperature variations of some basic thermodynamic quantities (such as entropy, specific heat, etc.) that might exhibit a rather spectacular thermal dependences especially near the zero-temperature 
phase transitions. Furthermore, several interesting extensions and generalizations of the present 
version of the Ising-Heisenberg model come into question besides the most obvious ones mentioned 
at the end of Section \ref{sec2}. For instance, it is possible to extend the present Ising-Heisenberg model by including higher-order triplet, quintuplet, and sextuplet interactions between the Ising and Heisenberg spins, or to solve exactly the analogous Ising-Heisenberg model with the Heisenberg spins $S>1/2$ that accounts for another interaction terms like the single-ion anisotropy and the biquadratic interaction. In this direction will continue our next work. 

\begin{acknowledgements}
This work was supported by the Slovak Research and Development Agency under the contract 
LPP-0107-06. The financial support provided by Ministry of Education of Slovak Republic 
under the grant No.~VEGA~1/0128/08 is also gratefully acknowledged.
\end{acknowledgements}

\end{document}